\documentstyle[12pt,nature,epsf]{article}

\begin{document}
\newcommand{\be}{\begin{eqnarray}}
\newcommand{\ee}{\end{eqnarray}}
\newcommand{\etal}{{\it{et al.}}}
\newcommand{\smass}{M_{\odot}}
\newcommand{\br}{{\bf r}}
\newcommand{\bV}{{{\bf v}}}

\parindent 0pt

\baselineskip 18pt

%\shorttitle{}
%\shortauthors{Funato}
%
% Title
%
\title{
The formation of 
Kuiper-belt Binaries through Exchange Reactions
}

%
% Authors
%

\author{Yoko Funato$^1${\footnote{
{\tt funato@chianti.c.u-tokyo.ac.jp}}},
Junichiro Makino$^2$,
Piet Hut$^3$,\\
Eiichiro Kokubo$^4$ \& Daisuke Kinoshita$^5$\\
{\small
$^1$ General System Studies,
University of Tokyo,
Komaba, Meguro-ku, Tokyo 153, Japan}\\
{\small $^2$ Department of Astronomy, University of Tokyo, Hongo, Bunkyo-ku, Tokyo 113, Japan,}\\
{\small $^3$ 
Institute for Advanced Study,
Princeton, NJ 08540, USA}\\
{\small $^4$ National Astronomical Observatory, Osawa, Mitaka-shi, Tokyo 180, Japan}\\
{\small $^5$ National Central University, 
Chung-Li 32054 Taiwan}
}

\maketitle

\begin{abstract}

{\bf 
\noindent
Recent observations
%\cite{Burnes2002,Veillet2002,Margot2002a,KSNoll2002,Schaller2003,KSNoll2003}
$^{1-6}$
have revealed an unexpectedly high binary fraction among the
Trans-Neptunian Objects (TNOs) that populate the Kuiper-belt.  The
discovered binaries have four characteristics
they comprise a few percent of the TNOs, the mass ratio of their
components is close to unity, their internal orbits are highly
eccentric, and the orbits are more than 100 times wider than the
primary's radius.  In contrast, theories of binary asteroid formation
tend to produce close, circular binaries.  Therefore, a new approach
is required to explain the unique characteristics of the TNO binaries.
Two models have been proposed\cite{Weidenschilling2002,Goldreich2002}.
Both, however, require extreme assumptions on the size distribution of
TNOs.  Here we show a mechanism which is guaranteed to produces
binaries of the required type during the early TNO growth phase, based
on only one plausible assumption, namely that initially TNOs were
formed through gravitational instabilities\cite{GoldreichWard1973} of
the protoplanetary dust layer.}

\end{abstract}

\def\simlt{\hbox{ \rlap{\raise 0.425ex\hbox{$<$}}\lower 0.65ex
  \hbox{$\sim$} }}
\def\simgt{\hbox{ \rlap{\raise 0.425ex\hbox{$>$}}\lower 0.65ex
  \hbox{$\sim$} }} 

The TNO binary with best known orbital elements, 1998WW31, has a mass
ratio $m_2/m_1 \sim 0.7$, eccentricity $e \sim 0.8$, and semi-major
axis $a \sim 2\times 10^4$ km\cite{Veillet2002} .  The inferred radii
of the components are $r_1 \sim 1.1r_2 \sim 10^2$ km, hence $a/r_1 >
10^2$, in stark contrast to main belt asteroid binaries where $m_2/m_1
\ll 1$, $e \sim 0$, and $a/r_1 \simlt 10$.  Asteroid binaries in the
main belt are probably formed by collisions\cite{Merline2003}, as in
the leading theory for the formation of the
Moon\cite{HartmannDavis1975,Kokubo2000}, which naturally explains the
extreme mass ratios and tight near-circular orbits\cite{Durda2001,
Durda2003}.

In order to find a completely different process for TNO binaries,
we can borrow ideas from dynamical binary formation in star clusters,
where many roughly equal-mass binaries are found.  There, we know of 
three formation paths: 1) two-body dissipational encounters\cite{FPR1975}; 2)
three-body binary formation\cite{Heggie1975}; and 3) exchange
reactions\cite{Heggie1975}. Figure 1 schematically shows these paths.

In the case of TNOs and asteroids, path 1 corresponds to the
standard theory for binary formation, namely tidal disruption and
giant impact during a close encounter ({\bf a} and {\bf b} in Figure
1).  They indeed occur: each TNO has grown through accretion, and much
of accretion has happened through collisions with an object comparable
in mass to that of the growing TNO itself\cite{KokuboIda1997,
Makino1998}.

Path 2 ({\bf c} in Figure 1) would require a near-simultaneous
encounter of three massive objects with low enough velocities to allow
an appreciable chance to leave two of the objects bound.  For this to
work, the random velocities of the most massive objects should be
significantly lower than their Hill velocities, $v_{{\rm H}} \equiv
R_{{\rm H}} \Omega(R)$, where $R_{{\rm H}}$ is the Hill radius of the
asteroid at a distance $R$ from the Sun and $\Omega(R)$ is its orbital
angular velocity.  This path could work if there are $\sim10^5$ 100
km--sized objects embedded in a sea of small ($<1 {\rm km}$) objects
\cite{Goldreich2002}.  This assumption, however, is at odds with
Goldreich and Ward's theory for the formation of
planetesimals\cite{GoldreichWard1973} through gravitational
instability, and it is hard to see how objects in the Kuiper-belt
could form from non-gravitational coagulation, because the time scales
are far too long\cite{Wetherill1990}.  In contrast, the gravitational
instability theory predicts the size of the initial bodies to be
$10-100 {\rm km}$.  Starting with these bodies would make path 2
ineffective, because the velocity dispersion would be higher than the
Hill velocity\cite{KokuboIda1997,KenyonLuu1998}.

A combination of paths 1 and 2 has been proposed
\cite{Weidenschilling2002}.  It was studied how a third massive body
could capture a collision product of two massive bodies if the third
body were near enough during the time of the collision.  This
mechanism seems unlikely to work, since it requires a number density
of massive objects about two orders of magnitude higher than the value
consistent with observations\cite{Goldreich2002}.

Another mechanism based on the dynamical friction from smaller bodies
that can let a hyperbolic encounter between two massive bodies produce
a bound orbit has been proposed\cite{Goldreich2002}.  As we mentioned
above, the gravitational instability theory for the formation of
planetesimals\cite{GoldreichWard1973} would exclude the existence of
such a sea of small objects.

Path 3 ({\bf d} in Figure 1) can operate on the binaries formed
through path 1 ({\bf b} in Figure 1), so we should check whether paths
1 and 3 together produce the right binaries in the right numbers.

Consider a relatively massive TNO primary with mass $m_1$, in a binary with
a secondary with mass $m_2 \ll m_1$.  If the binary encounters another
body with mass $m_3 \sim m_1$, the most likely result is an exchange
reaction, in which the incoming object replaces the original
secondary\cite{Spitzer1987}. Figure 2 shows an example of such a
reaction.

If the incoming velocity of the third body is small, the binding
energy of the resulting binary is comparable to or larger than that of the
initial binary, since otherwise the secondary could not escape. In
practice, the change in binding energy is relatively small, as
will be shown later.  Hence $m_1m_2/a_0 \approx m_{1}^{2}/a$ where $a$ is
the new semi-major axis after the exchange.  This implies $a/a_0
\approx m_1/m_2 \gg 1$.

For the exchange reaction to occur, the third body should approach the
binary to a distance comparable to $a_0$.  The angular momentum
carried away by the light body is small. Therefore, the periastron
distance of the resulted binary is comparable to that of the initial
parabolic orbit, in other words, to $a_0$. Thus, the eccentricity is
$e \sim 1-a_0/a \sim 1-m_2/m_1$.

We have run a series of numerical scattering experiments to obtain
cross sections for different outcomes of binary-single body
interactions (see Methods for details).  The results show that 80\% of
%%the total cross section of 
the cross sections corresponding to the change of the composition of the
binary is accounted for by processes leading to the formation
of binaries with two massive components.

In figure 3 the distribution for the semi-major axis is strongly
peaked around $a \sim 20$.  This result is in good agreement with the
simple argument presented above.  This peak arises from the exchanges
which take place at the first binary-single body encounter. The broad
peak around $a\sim 10$ arises from ``resonant encounters'', in which
encounters of two objects occur more than two times\cite{Hut1984}.
The eccentricity peaks at 0.95, as expected.  In the case of 1998WW31
with $r_1=75 {\rm km}$, our length unit corresponds to 1500 km.  In
figures 2 and 3 we attach this physical scale to give a concrete idea
of the size.  Figure 4 clearly shows that the orbital elements of
1998WW31 are consistent with the binary having formed through the
processes modeled here.

We now confront our second task: to check whether the exchange path
is efficient enough to produce the observed binaries.  Starting with
TNOs formed through gravitational instability of the protoplanetary dust
layer\cite{GoldreichWard1973}, the heaviest TNOs will accrete mass
primarily through collisions with TNOs of comparable
mass\cite{KokuboIda1997, Makino1998}.  Many of these collisions are of
the ``giant impact'' type, resulting in  tight strongly unequal-mass
binaries ({\bf b} in Figure 1).  Let us estimate how many such
binaries are formed and what fraction of them are converted to wide
binaries through exchange.

We assume that one in three collisions between comparable TNOs gives
rise to a binary\cite{Durda2003}.  The numerical simulations of
planetesimal collisions yield a sizable formation rate of satellites
of 43.3 \%\cite{Durda2003} .  Since the relative velocity between
objects in the TNO region is significantly lower than that assumed in
their simulations, the corresponding binary formation rate in the TNO
region is even higher.  To stay on the conservative side, however, we
have assumed a ratio of only 1/3.

In 2/3 of the cases when no binary is produced, we have to wait for a
typical time $T$ until another collision occurs.  When a binary is
formed, the cross section for subsequent interactions with a third
body, taking into account gravitational focusing, is in proportion to
$a_0$.  Therefore, our newly-formed binary will undergo an exchange
reaction on a time scale $(r/a_0)T \ll T$, leading to a significant
increase in $a$.  Strong three-body interactions will subsequently
occur on an shorter time scale $(r/a)T \ll T$.  As a result, the
semi-major axis will shrink systematically, with a `thermal'
distribution $f(e) = 2e$ favoring high eccentricity\cite{Heggie1975}.
When the orbit becomes small enough ($r/a \sim 0.03$), the change for
collisions in resonant encounters becomes
significant\cite{HutInagaki1985}.

Let us assume that an exchange reaction turns a ``giant impact'' binary
into a binary with $a \sim 300r$.  Each subsequent strong encounter
will on average shrink $a$ by a factor\cite{HH2003} $\sim 1.2$.
After a dozen encounters, $a \sim 30r$ and a collision is likely to
occur.  The time scale for each encounter to occur is $\sim (r/a)T$,
and the waiting time for the last encounter in this series is
$(1/30)T$.  Summing this series, we get a total waiting time of
$(T/30)/(1 - (1/1.2)) = 0.2T$ before a collision.  If three bodies
collide, or if the third body escapes, the resulting system may be a
single body (with an assumed chance of 2/3) or a strongly unequal-mass
binary (a chance of 1/3).  If the merger product and the third body are still bound,
we have an equal-mass wide binary.

Under these assumptions, in 1/3 of the cases, we wind up with an
equal-mass TNO binary with the observed properties for a period $\sim
0.2T$, compared to a 2/3 chance to wind up with a single TNO for a
period $\sim T$. Denoting by $N_S$ and $N_B$ the number of single
bodies and the number of binaries, respectively, we have
\begin{eqnarray}
\frac{dN_B}{dt} &=& \frac{1}{3}N_S \, -
                    \, \frac{1}{0.2} \,\frac{2}{3}N_B\nonumber\\
\frac{dN_S}{dt} &=& -\,\frac{dN_B}{dt}\nonumber
\end{eqnarray}
if we measure time in units of $T$. For the stationary state we have
$dN_B/dt = dN_S/dt=0$, and $N_B=0.1N_S$. Therefore, the predicted
binary fraction is $\sim 10$\%.  When accretion in the Kuiper-belt
region diminished, the number of single and binary objects was frozen,
with a ratio similar to this steady-state value.

While our arguments are  approximate, it is clear that after
cessation of the accretion stage at least several percent  of
the TNOs were accidentally left in such a binary phase.  The fact that
more than 1\% of the known TNOs are found to be in wide roughly
equal-mass binaries is thus a natural consequence of {\it any}
accretion model {\it independent of the assumed parameters} for the
density and velocity dispersion of the protoplanetary disk or the
duration of the accretion phase.  
We predict that future discoveries of TNO binaries will similarly show
roughly equal masses, large separations, and high eccentricities.

%\newpage

\section*{Methods}

%
%method --- version 2003/11/01
%

{\bf Numerical scattering experiments}
We performed 775,541 runs of numerical integrations of
binary-single body encounters.
We chose the relative velocity $v_{\infty}$ between
the intruder (single body) and the binary at infinity to be much
smaller than the orbital velocity of the binary components. In this
regime, the scattering cross section $\sigma$ is expressed as $\sigma
\propto
\sigma_0/v_{\infty}^2$, because of gravitational
focusing\cite{Weidenschilling2002,HH2003}. The parabolic orbit is the
limiting case of $v_{\infty}\rightarrow 0$ where the nominal cross section $\sigma$
diverges, but $\sigma_0$ converges to a finite value.

We fixed the initial orbital parameters
of the binary, and changed the relative orbit of the intruder to
obtain scattering cross sections. For encounters with finite
$v_{\infty}$, one should randomly change the direction of $v_{\infty}$
and also randomly select impact parameter vectors uniformly within a
circle of given radius. This radius should be large enough so that
encounters with impact parameter larger than this radius would not
result in exchange or collision. For parabolic encounters, this
procedure choice corresponds to choosing the periastron distance $r_{p}$ from
a uniform distribution with an upper limit and selecting random orientations for
the orbital plane and direction of the periastron. As an upper limit,
we chose $20a_0$, where $a_0$ is the semi-major axis of the initial
binary.  Initial separations between the intruder and the binary were
$100a_0$.

We use units in which $G=m_1=m_3=a_0=1$, where $G$ is the
gravitational constant, $m_1$ and $m_3$ are the masses of the heavier
body of the initial binary and the intruder. The mass of the secondary
component is $m_2=0.05$.  The radii of the bodies are $r_1=r_3=0.05$ and
$r_2=r_1(m_2/m_1)^{1/3}\approx 0.01842$. The initial binary is circular.
These values are typical for main-belt binary asteroids, with $m_2/m_1
< 0.1$, and separations are $5-40$ times the radius of the primary.

We terminated the numerical integration when (a) one of the three
bodies escaped beyond the Hill radius, or (b) when two bodies physically
collided. For (a) we chose a Hill radius of $r_{\rm H}=100a_0$, since a
typical value for a TNO with radius $r$, $m = 10^{21}$ g and $R = 
40 $ AU is $ a = 1000 r$, and the typical
semi-major axis of a close, circular planetesimal binary is $a_0 = 10r$.
When a collision occurred, the collision product was assumed to
inherit the mass and motion of the center of mass of the two bodies. We
adapted the radius so as to conserve the density. In both cases, we
calculated the orbit of the remaining two bodies to determine the final
state.

We integrated the equations of motion of the three bodies in Cartesian
coordinates. Integration was done with an 8th order time-symmetric
variable-stepsize Hermite method.  To cross-check the results, we also
performed two other sets of scattering experiments, one using the
scattering package of STARLAB\cite{McMH1996}, and the other using a
4th order Hermite scheme.  In all cases, we used completely different
programs for generating initial conditions, as well as for orbit
integration and for the reduction of the results.  All three
experiments gave the same outcome, within the expected fluctuations
associated with the finite number of runs.

There are six possible configuration changes that can take place as a
result of a scattering event:
(a) an exchange reaction resulting in a massive--massive binary;
(b) an exchange reaction resulting in a massive--light binary;
(c) a merger of bodies 1 and 2 resulting in a massive--massive binary;
(d) a merger resulting in a twice-as-massive--light binary;
(e) a merger of bodies 2 and 3 resulting in a massive--massive binary;
(f) no binary is left, after three-body merging or two-body merging
followed by escape.

The cross section $\sigma_0$ for each event is as follows: (a) 12.1
(b) 1.3 (c) 0.9 (d) 1.3 (e) 0.9 (f) 1.2 in our units.  For our
scattering experiments with $r_p< 20 a_0$, we found a cross section
for preservation of the original binary of 1243.9 in our units, which
implies that in 99 \% of our experiments no configuration change took place.
We also found that for $r_p> 20 a_0$ no merger or exchange occurred, so
our cross section measurements are complete.

Another logical possibility would be: (g) all three objects become
unbound.  However, for that to occur the total energy would have to be
positive.  In our case, the low relative velocity of the incoming
third body guarantees the total energy to be negative, hence (g)
cannot occur.

The physical reason for the dominance of (a) is the statistical
tendency for equipartition of kinetic energy, which implies much
higher velocities for much lighter bodies and therefore a much larger
chance for those lighter bodies to escape.

We acknowledge helpful comments on our manuscript by Peter Goldreich,
Roman Rafikov, Re'em Sari, Keith S. Noll and Dan Durda.

\newpage

\begin{figure}
\epsfxsize 18cm 
\epsffile{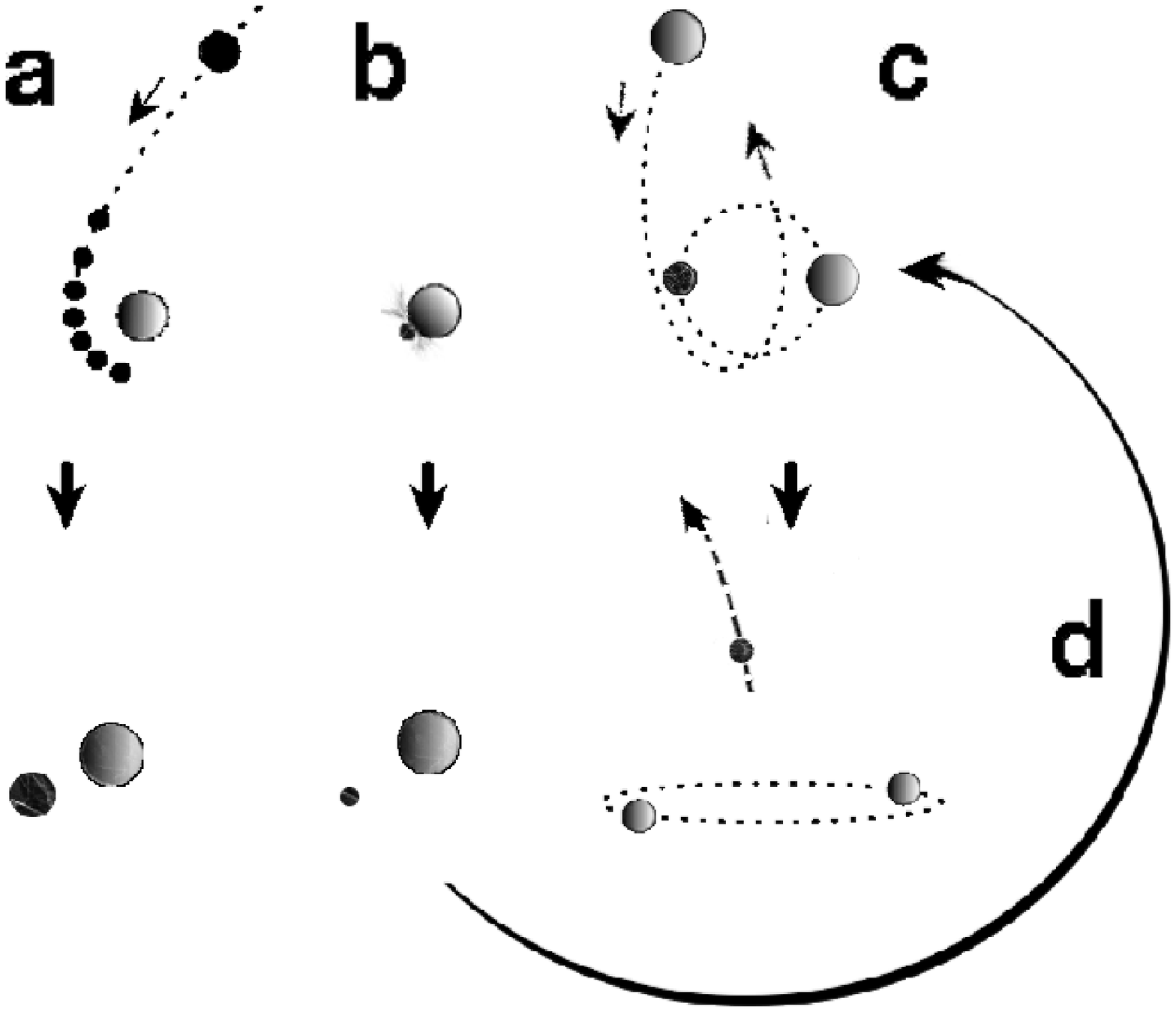}
\caption{
Paths of formation of binaries. Path
({\bf a}) is the formation through tidal disruption of one object followed
by coagulation of fragments during a close encounter with the other; ({\bf b}) a
giant impact, where collision debris coagulates into a ``moon''; ({\bf c})
an exchange reaction, where the incoming body replaces one component
of a binary; ({\bf d}) a combination of ({\bf b}) and ({\bf c}).}
\end{figure}

\begin{figure}
\epsfxsize 18cm 
\epsffile{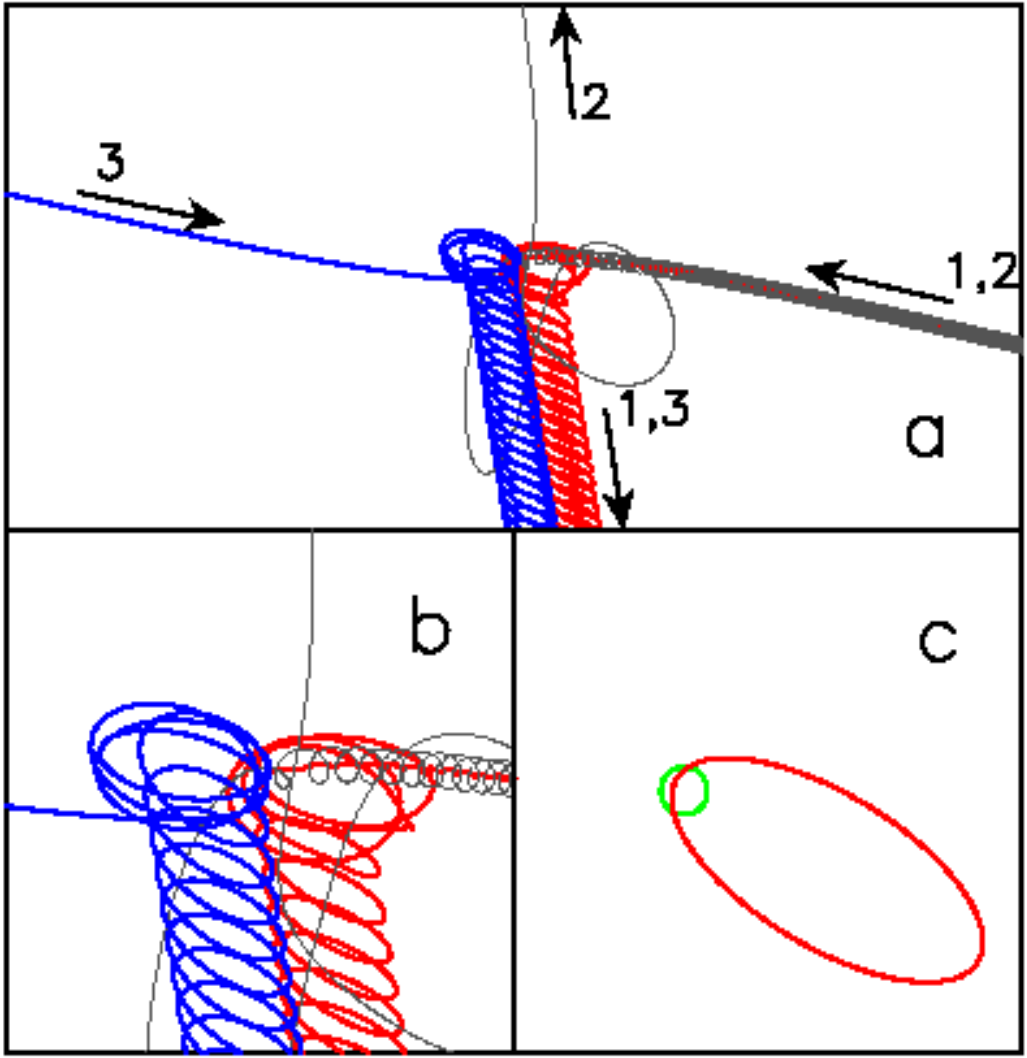}
\caption{
An example of a binary--single-body exchange
interaction.  Bodies 1 and 2 having masses $m_1=1$ and $m_2=0.1$,
respectively, form a binary with an initially circular orbit.  Body
3, with mass $m_3=1$, encounters the binary on an initially parabolic
orbit.  In panel {\bf a}, the whole scattering process is shown.
Panel {\bf b} shows the complex central interaction in more detail,
while panel {\bf c} shows the orbits of the initial (green) and final
(red) binary, respectively.  The final binary orbit is highly
eccentric and much wider than the initial circular binary orbit.
}
\end{figure}

\begin{figure}[t]
\epsfxsize 18cm 
\epsffile{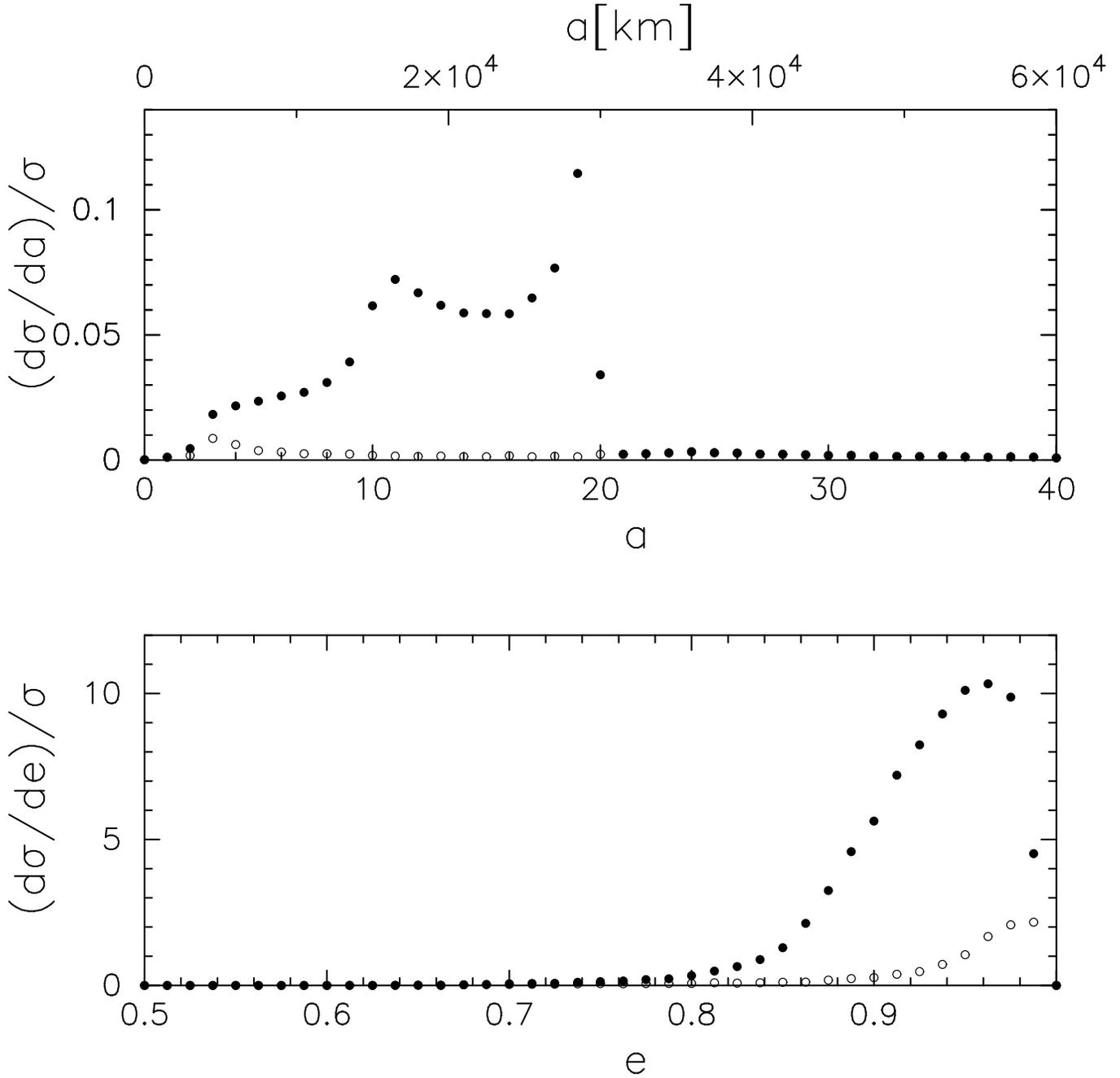}
\caption{
Normalized differential cross sections for the formation of a
`massive-massive' binary.  Normalized differential cross sections for
the event-types (a), (c) and (e) in Methods are plotted with respect
to the semi-major axis $a$ (top panel), and eccentricity $e$ (bottom
panel) of the final binary.  The physical units are given for
reference at the top of the figure.  The filled points are the total
values, while the open circles are the contributions from the merger
events (types (c) and (e) in Methods).
}
\end{figure}

\begin{figure}[b]
\epsfxsize 18cm 
\epsffile{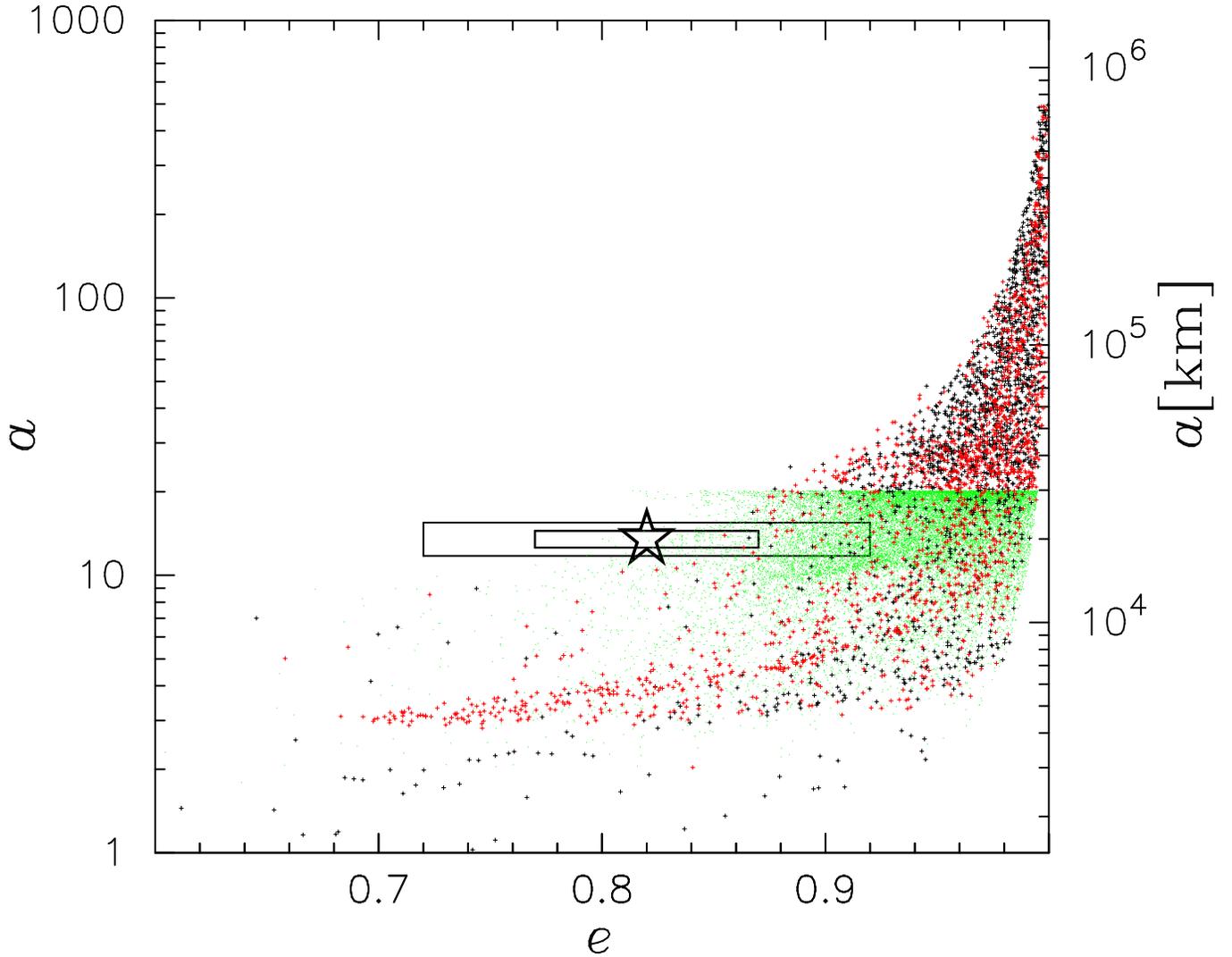}
\caption{ 
Distribution of orbital properties of `massive-massive' binaries
formed in our scattering experiments. Here $a$ and $e$ have the same
meanings as in fig. 3.  Contributions from exchange reactions,
event-type (a) in Methods, are shown by green dots. They are limited
by energy conservation to $a \simlt 20$, and give rise to the
horizontal rim in the middle of the figure.  Contributions involving
mergers, types (c, red dots) and (e, black dots), can lead to $a$
values all the way to the Hill radius $a \approx 10^3$.  The star
symbol shows the observed orbit for 1998WW31.  Boxes around the star
indicate the observational 1- and 2-$\sigma$ error bars.
}
\end{figure}

\end{document}